# Integrated Photonic Quantum Technologies

Jianwei Wang[1], Fabio Sciarrino[2], Anthony Laing[3], Mark G. Thompson[3]

1. State Key Laboratory for Mesoscopic Physics and Collaborative Innovation Center of Quantum Matter, School of Physics, Peking University, Beijing 100871, China
2. Dipartimento di Fisica - Sapienza Università di Roma, P.le Aldo Moro 5, 00185, Roma, Italy
3. Quantum Engineering Technology Labs, H. H. Wills Physics Laboratory and Department of Electrical and Electronic Engineering, University of Bristol, BS8 1FD, Bristol, United Kingdom

email: mark.thompson@bristol.ac.uk

**Generations of technologies with fundamentally new information processing capabilities will emerge if microscopic physical systems can be controlled to encode, transmit, and process quantum information, at scale and with high fidelity. In the decade after its 2008 inception, the technology of integrated quantum photonics enabled the generation, processing, and detection of quantum states of light, at a steadily increasing scale and level of complexity. Using both established and advanced fabrication techniques, the field progressed from the demonstrations of fixed circuits comprising few components and operating on two photons, to programmable circuitry approaching 1000 components with integrated generation of multi-photon states. A continuation in this trend over the next decade would usher in a versatile platform for future quantum technologies. This Review summarises the advances in integrated photonic quantum technologies (materials, devices, and functionality), and its demonstrated on-chip applications including secure quantum communications, simulations of quantum physical and chemical systems, Boson sampling, and linear-optic quantum information processing.**

## Introduction

Pioneering investigations of quantum science, including tests of entanglement [1], generation of squeezed light [2], demonstrations of quantum teleportation [3], and loophole-free tests of Bell nonlocality [4], were possible because photons are excellent, low-noise carriers of quantum information. Weakly coupled to their environment, photons do not suffer the decoherence issues of matter-based systems, and thus do not require operation at *mK* temperatures or in high vacuum. The translation of quantum information science to real world technologies promises extreme advantages for certain tasks in communication [5], computation [6], and simulation [7]. However, implementing large instances of quantum algorithms or the deployment of a global quantum communication network, will require new levels of sophistication in the manufacture of quantum technologies with a large number of components. Exploiting single particles of light as quantum information carriers, together with wafer-scale fabrication, *Integrated Quantum Photonics* (IQP) is a compelling platform for the future of quantum technologies.

While the photon-photon interactions required to realise quantum logic can be mediated through light-matter interactions, in systems including quantum dots and colour-centers [8], routes to scaling such systems remain challenging. In 2001, a theoretical breakthrough presented a scheme for universal quantum computation using photons in linear optics, in which photons interact non-deterministically via quantum interference and measurement-induced nonlinearities [9, 10] – requiring no light-matter interactions. In the same year, a measurement-based quantum computation (MBQC) scheme was proposed enabling a significantly more resource-efficient quantum computing model, that is particularly well suited to IQP [11, 12]. While realising a universal fault-tolerant quantum computer remains challenging in any system at present, IQP provides a potentially low-resource route toward a specific but powerful quantum machine that could outperform a classical computer in a specific task, referred to as Boson sampling [13]. Photons provide the only viable technology to connect distributed quantum systems, or for quantum communication systems [5].

In contrast to the well-established platforms of trapped ion and superconducting systems (their first logic operation were demonstrated in the 90s), IQP is relatively new, with the first integrated photonic quantum logic gate demonstrated in 2008 [14]. Yet, despite its relative immaturity, IQP can leverage commercially available tools from the complementary metal-oxide-semiconductor (CMOS) industry. Using optical waveguides to guide and route single-photons, providing phase-stable and miniaturisable circuitry, core functionalities of entangled state generation, manipulation and detection have been demonstrated. While photonic experiments in bulk-optics continue to deliver significant results [15], large-scale quantum information processing with hundreds or thousands of photons will almost certainly require the technological functionalities and scaling that only integrated photonics can provide.

In this article, we review the rapid progress that has been made over the last decade in IQP technology and applications. We first review a range of integrated photonics platforms, and recent progress in integrated single-photon sources (SPS), linear-optic quantum circuits, and integrated single-photon detectors (SPD). We focus on applications of IQP, including chip-scale quantum communications, quantum information processing and computing, Boson sampling, and quantum simulation of physical and chemical systems. Finally, we discuss the challenges and opportunities of realising large-scale, monolithic IQP circuits for quantum applications.

### Integrated quantum photonic platforms and key devices

IQP provides a versatile platform for on-chip generation, processing, and detection of quantum states of light. A landmark result in this field is a demonstration of quantum interference and controlled-NOT entangling gate in silica-on-insulator optical waveguide circuits in 2008 (**Fig.2a**) [14]. The achieved high fidelity demonstrates the inherent stability and controllability of the IQP platform. On-chip state-preparation and manipulation of single-qubit and two-qubit states were also implemented in this silica platform [16–18], and in the laser-writing silica platform (**Fig.2b**) [19–22]. In the last decade, significant progress has been made in materials, devices, functionality, and implementations of quantum protocols and algorithms. **Figure 1** summarises key milestones in the field, which lay the groundwork for the development of the IQP technology and its applications.

**IQP platforms.** A range of optical waveguide systems have been developed for IQP, including laser-writing silica [19–22], silica-on-insulator ($SiO_2$) [14, 16–18], silicon-on-insulator (Si) [23–26], silicon nitride ($Si_3N_4$) [27–30], lithium niobate (LN) [31–33], gallium arsenide (GaAs) [34–36], indium phosphide (InP) [37, 38], and others. We refer detailed discussions on the IQP platforms in other reviews, while here highlight a few of them. For example, silicon waveguides can tightly confine light, allowing single-photon generations in waveguides and high-density integration (**Fig.2c,i**) [39]. $Si_3N_4$ can generate single-photons over a broadband window [27]. Laser-writing 3D circuits can provide possibilities for studying complex physical Hamiltonians (**Fig.2b**) [40]. Whilst the above example are all passive materials, LN [32], GaAs [35] and InP [37] own large electro-optical effects that allow fast manipulation of single-photon states.

**Qubit encoding and manipulation.** An on-chip toolbox has been developed to encode, transmit, and process quantum information in various degree-of-freedoms (DoFs) of single-photons (**Box 1**). Flexibly in controlling the photon's DoFs greatly enriches functionalities. For example, encoding states in time and polarisation enables robust chip-to-chip quantum interconnection via fiber or free-space. Encoding qubits in path allows extremely low-error single-qubit and two-qubit operations (**Box 2**), key for on-chip quantum information processing. Si-based MZIs with a 65dB on-off ratio has been achieved with near-perfect beamsplitters [41], equivalent to having a Pauli-Z error rate of $< 10^{-6}$ [42]. With ultra-high precise operations and measurements, the overhead resources for MBQC can be significantly reduced. The high levels of controllability of photonic states has also been confirmed by the demonstrations of two-photon quantum interference with a near-unity visibility of 100.1±0.4% in $SiO_2$ [18] and 100.0±0.4% in Si platforms [25].

**Integrated single-photon sources.** Harnessing the power of IQP requires on-chip generation of a large number of identical single-photons. We discuss two types of integrated SPSs: parametric photon-pair source and quantum dot (QD) single-photon source.

Pumping nonlinear optical waveguides or cavities allows the generation of photon-pairs, via the spontaneous four-wave mixing (SFWM) (**Fig.2c**) or spontaneous parametric down-conversion (SPDC) process (**Fig.2d**). These SPSs have been reported in $\chi^3$ waveguides (e.g., Si [46], $SiO_2$ [47] and $Si_3N_4$ [27]), and in $\chi^2$ waveguides (e.g., GaAs [34] and LN [32]). A notable feature of integrated parametric SPSs is that they can be engineered into arrays are highly identical sources, each one individually controllable. We give two state-of-the-art examples: an array of 18 SFWM-SPSs in $SiO_2$-waveguides [47] (4 SFWM-SPSs in Si-microresonators [48]) can produce heralded single-photons with a 52% (50%) heralding efficiency and 95% (91%) photon indistinguishability from separated SPSs. An issue for parametric SPSs however is that photons are produced non-deterministically with a 5%-10% probability. This can be improved by exploiting the multiplexing technique, such as in time [49] and spatial [50] domains. In Ref. [49], a state-of-the-art multiplexing source is demonstrated in bulk-optics, achieving a 66.7% high probability of single-photons collected into a fiber and high indistinguishability of ~90%.

QD-SPSs promise deterministic generation of single-photons, and particularly self-assembled InGaAs/GaAs QD-SPSs hold the best performance [51]. A breakthrough in 2013 demonstrated near-optimal single-photon emissions from a single QD, by using a resonant excitation technique [52]. Photons with a 99.1% (99.7%) single-photon purity, 66% (65%) extraction efficiency, and 98.5% (99.6%) indistinguishability have been produced in a single QD [53] (in several QDs samples [54]). This type of QD-SPSs in micropillar emit photons out-of-plane (**Fig.2e**), presenting ease of fiber-coupling but difficulty in waveguide integration. Instead, QD-SPSs in photonic-crystal waveguides allow near-unity preferential emission into the waveguide [55]. A major challenge however is to create multiple, identical QD-SPSs, due to the difficulty in reproducing the samples. A solution is to actively de-multiplex the single-photons from a single dot into different spatial modes [56]. This scheme produces multiple photons at the loss of overall rate, while having shown a great success in recent Boson sampling implementations [57–59].

**Integrated single-photon detectors.** On-chip detection of single-photons allows the ultimate readout of quantum information. Nowadays several SPD technologies are available [60], e.g., avalanche photodiodes, superconducting nanowire SPD (SNSPD), and transition edge sensor (TES). Fully integrated SNSPDs have been patterned atop GaAs [36], Si [26], and $Si_3N_4$ [30] waveguides. A breakthrough in 2012 showed evanescently-coupled Si-waveguide SNSPDs with a 91% detection efficiency, 18ps jitter, and 50Hz dark count (**Fig.2g**) [26]. Instead of direct deposition, SNSPDs fabricated on a $Si_3N_4$ membrane can be flexibly transferred to other substrates (**Fig.2j**) [61]. Moreover, photon-number-resolving (PNR) detection is required in many quantum protocols [60]. The PNR TESs have been evanescently integrated on $SiO_2$ (**Fig.2f**) [62] and LN waveguides [33], that is able to resolve up to 5 photons. A series of integrated SNSPDs atop GaAs waveguide also allows the on-chip PNR detection of up to 4 photons [63].

## Chip-based quantum communications

Quantum communications aims to share encrypted keys between two parties, with a security based on the law of quantum mechanics [5]. The best-known example is quantum key distribution (QKD). Integrated photonics enables a robust, miniaturized, and low-cost platform to realise QKD transmitter and receiver devices. We review recent experimental progress in chip-based QKD hardware that utilises superposition-based and entanglement-based protocols. Discussions on advanced QKD protocols and physical realisations in bulk-optics and fiber-optics are referred to in other reviews [5].

**Superposition-based QKD.** Integrated photonics is well established in the telecommunication industry, that trancieves data in the global scale. It is also natural to use integrated photonics for practical QKD to trancieve encrypted keys. Pioneer works began in 2004: $SiO_2$-based optical interferometers were fabricated for time-bin QKD systems in fiber [64].

The first fully integrated chip-to-chip QKD system was implemented with an InP transmitter and a silicon oxynitride ($SiO_xN_y$) receiver (**Fig.3a**) [37]. The InP transmitter incorporates all the necessary components including a tuable laser, attenuator, and electro-optic phase modulators. The $SiO_xN_y$ receiver consists of thermo-optic phase shifters, allowing for digitally reconfigurable delay line and state measurement. Photons were detected by off-chip SNSPDs. These transmitter and receiver devices provide a complete chip-to-chip QKD solution, and could be programmed to implement multiple protocols, including: BB84 at a 560MHz state rate; coherent-one-way operating at a 860MHz state rate; differential-phase-shift at a 1.76GHz state rate [37].

Silicon-photonics is appealing for QKD applications, due to its compatibility with CMOS process and telecommunication infrastructure. Fast manipulation of photons in Si however is challenging, due to the lack of efficient electro-optic pure-phase modulation. However, carrier injection and depletion modulations, can be adopted for state preparation for QKD typically at a rate of MHz-GHz. Recently, three groups demonstrated Si-based QKD transmitters: in ref. [65], the transmitter prepares polarisation-encoded qubits for a BB84 system over a 5km fiber (**Fig.3c**); in ref. [66], two transmitters were developed, one for polarisation-encoded and the other for time-bin-encoded BB84 systems; in ref. [67], the transmitter was tested in real-life QKD system, in which Alice was located in Cambridge and Bob in Lexington (**Fig.3d**). In these demonstrations, secret key rates of kbps-Mbps and low quantum bit error rates of 1.0%-5.4% were obtained, and all implementations used external laser sources and weak coherent photon states.

**Multiplexing QKD.** Secreted key rates can be further increased by exploiting multiplexing techniques. Take wavelength-division multiplexing (WDM) as an example, multiple keys can be co-distributed in a single fiber but at different wavelength channels. A proof-of-concept demonstration of WDM-QKD system that consists of two InP transmitters and a $SiO_xN_y$ receiver, shows an increase of key rates by a factor of two, to 1.11Mbit/s in a 20km emulated fiber [68]. Another example is an implementation of a multi-dimensional chip-to-chip QKD system on two Si-chips, based on space-division multiplexing in a multicore fiber [69]. The prepare-and-measure protocol is applied for four-dimensional states by the Si transmitter and receiver. Low quantum bit error was achieved in this experiment.

**MDI-QKD.** QKD is rigorously secured, however its practically implementations can be imperfect, leaving loopholes that undermine its security. One significant example is detector side attacks, and measurement-device-independent (MDI) QKD was proposed and demonstrated to tackle such vulnerabilities. Recently, two independent demonstrations tested the feasibility of using integrated photonics for MDI-QKD by demonstrating Hong-Ou-Mandel (HOM) interference, a key aspect of the MDI protocol. HOM interference between weak coherent states from two independent InP chips was observed with a visibility of 46.5±0.8% in ref. [70], and from two Si/III-V integrated lasers with a high visibility of 46±2% in ref. [71] (note maximal achievable visibility is 50%). Integrated photonics may provide an ideal platform for multi-user chip-based MDI-QKD networks.

**Entanglement-based QKD.** Fully device-independent (DI) QKD and quantum internet rely on entanglement generation and distribution. In an early demonstration of on-chip photon generation, in 2002 a PPLN waveguide was used to create entangled time-bins for QKD [31] – an encoding scheme that provides inherent stability when transmitted in optical fiber. Recently, time-bin entangled states were also generated in GaAs [72] and $Si_3N_4$ [28] systems. Entangled time-bins across a visible-telecom range were produced in a delicately engineered $Si_3N_4$-microresonator and distributed over a 20km fibre [27]. The first chip-to-chip entanglement distribution [73] (**Fig.3b**), and recently quantum teleportation [48] were demonstrated between two programmable Si-chips, that integrate all necessary components including the entangled photon sources and Bell analyser. The systems were stabilised using a path-to-polarization conversion technique, that was also adopted in the prepare-and-measure QKD systems [65–67].

**Integrated QRNG.** Quantum random number generators (QRNG) provide a high bandwidth source of true random numbers with applications many fields, particularly in QKD. Integrated QRNGs have been demonstrated based on various non-deterministic quantum process, e.g., phase fluctuation [38, 74], vacuum fluctuation [75], and non-locality [44]. For example, fully integrated QRNGs have been demonstrated in InP-chips (**Fig.3e**) [38, 74], in which random numbers were generated from the random phases of two-laser interference. All these QRNGs operate in the Gbps regime. Quantum theory allows a stronger form of certified randomness, i.e., DI-QRNG. By violating the Bell inequalities for multi-dimensional entangled states in a Si-chip, randomness was certified in the fully DI scenario [44].

## On-chip quantum information processing with photons

In the gate-based quantum computing model, interactions between photons are realised by including ancillary circuitry and photons [9]. The partial state collapse, that follows from detecting the ancillary photons in the ancillary modes, implements the required quantum logic. By teleporting quantum information across photonic circuitry, from gates that operated successfully, quantum computation is possible [10]. Yet the overheads of the originally proposed scheme appeared to be forbidding at a practical level. The MBQC scheme presents a significantly more resource-efficient quantum computing model [11, 12]. The challenge from requiring deterministic gates can be shifted to constructing a generic entangled cluster-state, on which any quantum computation can be mapped by a sequence of measurements. Remarkably, MBQC offers tremendous advantages for photonic realisations because of its comparability with photon's probabilistic nature [12].

**Gate-based quantum information processing.** The building-blocks for the gate-based scheme have been demonstrated in IQP chips. For example, the CNOT gate was first demonstrated in $SiO_2$ [14], then in laser-writing silica [22], Si [39], and others. These initial demonstrations implemented a unheralded CNOT scheme that does not include the ancillary photons, nevertheless it was possible to demonstrate a rudimentary version of

Shor's factoring algorithm using two CNOT gates (**Fig.4a**) [76]. It allows the factorisation of 15 with a fidelity of 99±1%. The first integrated heralded CNOT gate was performed using a fully programmable $SiO_2$-chip that was able to implement universal linear-optic operations [77]. The type of teleportation required to shuttle successfully quantum information from heralded gates was also demonstrated on-chip with three photonic qubits and one CNOT gate in a laser-writing $SiO_2$-chip (**Fig.4b**) [78]. It achieved the single-chip teleportation of single-qubit states with an average fidelity of 89±3%.

**Programmable quantum chips for multifunctional information processing.** Re-programming photonic chips allows the processing of multiple quantum tasks and algorithms in a single chip. The first fully programmable two-qubit quantum photonic processor was demonstrated in a silica chip in 2011 [17]. The device includes single-qubit preparation and measurement, as well as two-qubit entangling operation. It was reconfigured to perform 1000s of experiments while remaining high fidelity, to study the wave-particle duality in a quantum delayed-choice experiment, and to implement a quantum algorithm that can compute the ground-state energy of molecules [79]. The first reconfigurable laser-writing device was also demonstrated recently [80]. In **Fig.4c**, a Si-photonic quantum processor able to initialize, operate and analyse arbitrary two-qubit states and processes, was demonstrated by adopting a linear-combination scheme [81]. The device was programmed to implement 98 different logic gates (e.g., CNOT, CZ, CH and SWAP), and achieved an average quantum process fidelity of ~93%. Multiple algorithms are implemented in the device, such as a quantum approximate optimization algorithm for a three-example constraint satisfaction problem.

Universal linear-optic circuits, able to implement all possible quantum protocols, have been proposed and experimental realised in programmable optical circuits. Its first realization was in a single $SiO_2$-chip in 2015 (**Fig.4d**) [77], which consists of a six-mode triangularly arranged MZI network. The versatility of this universal circuit was demonstrated for several key applications, including the implementations of heralded CNOT gate, Boson sampling circuits with verification tests, 6-dimensional complex Hadamard operations [77], and quantum simulation of vibrational dynamics of molecules [82]. Recently, an eight-mode universal linear-optic circuit was also demonstrated in a $Si_3N_4$-chip [83].

**Entanglement GMM and MBQC.** The generation, manipulation and measurement (GMM) of entangled states is at the heart of MBQC. On-chip GMM of entangled states requires an integration of qubit sources and circuits. For example, the first GMM of entangled two-qubit states encoded in dual-paths were demonstrated in a Si-chip, with the inclusion of ring-resonator-based SFWM photon sources and circuits [84]. The arbitrary GMM of 15-dimensional entanglement in multi-paths was demonstrated in a large-scale Si-chip with 16 waveguide SFWM photon sources [44]. A coherent excitation of multiple frequency-bin of two-photon results in multi-dimensional entanglement in the frequency domain, in which arbitrary operations on qudit can be performed using standard fiber-components [43].

MBQC relies on large entangled cluster states. The first on-chip GMM of four-qubit cluster state was realised in a laser-writing chip (**Fig.4e**), by manipulating both the polarisation and path DoFs of two photons [85]. Four-qubit linear-cluster and box-cluster states were generated and measured. Importantly, the Grover's search algorithm for a four-element database problem was demonstrated for the first time using photonic MBQC devices [85]. This implementation of multi-DoFs entanglement can be extended to multi-dimensional systems. For example, three-level four-qubit cluster states were generated in a hydex microresonator, by independently manipulating the time and frequency DoFs of two photons [86]. A proof-of-concept qutrit-MBQC was performed to show high noise robustness.

The on-chip GMM of four-photon four-qubit graph states was demonstrated in a Si-chip (**Fig. 4f**) [87]. The device includes four waveguide SFWM-SPSs creating two-pairs of photons and a reconfigurable two-qubit linear-optic operator. Performing the fusion operation or CZ operation (see **Box 2**) generates the line-graph or star-graph states. Recently, a state-of-the-art Si-photonic quantum device having an array of nearly-optimal FWM-SPSs in microresonators and high-quality reconfigurable circuits was coherently controlled to demonstrate and certificate the genuine multi-partite entanglement [48].

## On-chip sampling of photons and quantum simulation

Intermediate-scale quantum photonic devices offer the possibility of demonstrating a quantum advantage over classical computers for specific tasks [88]. Boson sampling is destined for such a task and highly suited to the IQP implementation. While classical computers are fundamentally inefficient at tackling complex systems govern by the law of quantum mechanics, engineering quantum devices that are inherently controllable to efficiently simulate quantum systems is a promising approach [7].

**Multiphoton interference.** The adoption of IQP has allowed the observation of the generalised HOM effect (**Box 2**). The observation of a three photon bosonic coalescence via a tritter device, the extension of the beamsplitter from two to three input modes, has been realised by exploiting laser-writing technique [89], via an integrated device containing three coupled interferometers [90].

**Boson sampling.** The Boson sampling problem consists of simulating the following quantum experiment: input $n$ bosons in different modes of an $m$-mode linear interferometer and sample events from the distribution of bosons at the interferometer's output modes [13]. If performed with indistinguishable bosons, this experiment results in an output distribution that is hard to sample, even approximately, on classical computers. In fact, the calculation of the probability associated with each observed Boson sampler event requires the estimation of a permanent, a notoriously intractable matrix function [13]. The input for the classical simulation consists of the unitary matrix $U$ describing the interferometer and the list of $n$ input modes used. It is desirable to choose $U$ randomly, both to avoid regularities that could simplify the classical simulation and because the main hardness-of-simulation argument holds only for uniformly sampled unitaries.

The first experimental demonstrations of Boson sampling were reported with $n = 3$ photons [91–95], mainly in integrated photonics. Since then, several investigations have been performed to study the scalability in imperfect conditions, e.g., in the presence of losses, partial distinguishability [96] and generic experimental errors. To date the highest number of photons generated in a Boson sampling experiment have been through the de-multiplexing of a single QD-SPS [57–59]: **Fig.5b**.

A generalised Boson sampling scheme, known as scattershot, has been proposed to overcome the limitation of probabilistic SPDC/SFWM SPSs. It has been demonstrated first on chip via bulk-optic SPDC-SPSs [15, 97] (**Fig.5a**), and recently via a fully integrated chip via SFWM-SPSs [98]. An alternative model known as Gaussian Boson sampling that uses squeezed light has been proposed and demonstrated in bulk-optics. Both Gaussian and scattershot Boson sampling have been implemented in the same Si-photonic chip [98].

**Validation of Boson sampling.** An open problem is to what extent the correctness of the outcomes of quantum hardware can be certified. In this framework, Boson sampling represents a relevant benchmark for testing different procedures to validate the obtained calculation/simulation: could it be possible to exclude that the Boson sampling device is sampling from another specific different probability distribution, instead of the nominal one? To which extent a validation procedure can be pushed towards a full certification of the device? A full certification of Boson sampling is believed to be not possible [99], accordingly all demonstrated protocols aim at ruling out the most plausible alternative scenarios. Currently, several protocols exist to validate Boson sampling, most of which have already been successfully demonstrated experimentally with integrated photonics hardware [77, 94, 100–102].

**Regime of quantum advantage.** A fundamental question is which physical resources in terms of number of photons, number of modes and other relevant physical parameters, are necessary to achieve the regime of quantum advantage. In the last few years an increasing effort has been devoted to define the classical limits on the simulation of Boson sampling: experiments will need to operate with >20 photons before they can start to challenge conventional computers (regime of quantum advantage) and >50 photons to surpass supercomputers (regime of quantum supremacy) [103].

**Simulation via quantum walks.** Quantum walks (QW) are the extension of classical random walks to a quantum framework and can be adopted as a resource structure for quantum simulations. Two classes of QWs exist: discrete-time QW, where the evolution is stroboscopic since it occurs in discrete steps; continuous-time QW, where the evolution on a given lattice is described by a Hamiltonian associated to the coupling between the different sites. QWs have been implemented both in the single-particle, and multi-particle regimes where quantum interference between the different particles is occurring during the quantum dynamics.

The discrete-time QW allowed the observation of single-particle and two-particle Anderson localization with bosonic and fermionic statistics (**Fig.5c**) [104, 105], and the entanglement growth after a spin chain quench [106]. The continuous-time QW has been adopted to identify signature of bosonic coalescence [107], to simulate Fano resonance [108], to shed light on the interplay between quantum coherence and noise for assist transport processes in complex networks [40, 41, 109] (**Fig.5d**), and to quantum fast hitting on hexagonal graphs [110]. The adoption of laser-writing technology has been adopted to engineer chip with 3D geometry leading to implementations of 2D QW for the simulation of different physical process [40, 111].

**Molecular simulation.** Efficient simulation of molecular properties is fundamentally interesting in chemistry [7]. Quantum phase estimation [112] and variational eigensolver [79] were implemented in IQP chips to compute the ground-state energy of molecules (**Fig.5e**). An algorithm combining these two approaches was implemented to approximate eigenvalues for both ground and excited states [113]. Moreover, a modification of Boson sampling allows molecular vibrionic spectra to be calculated with the inclusion of squeezed photons [114], and an alternative scheme was proposed and realised in a universal linear-optic device (**Fig.5f**) [82]. The question is how can underpinning models of the simulated systems be validated? Using a quantum Hamiltonian learning method, the fidelity of the simulation to the actual physical system can be measured by the consistency of the predictions with the obtained data. This algorithm was verified in a Si-photonic simulator (**Fig.5g**) [115].

## Challenges and outlook

Silicon photonics has emerged as a particularly exceptional platform for IQP, mainly due to its availability, maturity, natural integration of SPSs, and ability to realise high-yield large-scale optical circuits. In **Fig.6**, we summarise the scaling of integrated quantum photonic circuits in silicon: from a single MZI adopted for the on-chip HOM-interference experiment in 2012 [24], to a state-of-the-art large-scale quantum photonic device having 671 components that was used for the GMM of multi-dimensional entanglement [44]. The information processing capabilities have been significantly enhanced through access to such complex quantum devices. We expect this trend of increased circuit complexity to continue, and could envisage future very-large-scale integration of quantum photonic circuits containing millions of discrete components. Our optimism comes from the fact that manufacturing wafer-scale photonic circuits in silicon is compatible with "zero-change" of current CMOS fabrication process. Important examples from classical photonics include 16,000 component phased-array [116] and a demonstration of wafer-scale silicon photonic switches beyond die size limit, integrating 57,600 individual switches [117].

Increasing the number of photon states is critical to further enhancing the capabilities of the IQP technology platform. State-of-the-art IQP devices have demonstrated the generation and processing of up to 8-photon states in silicon-photonics [98]. However, processing significantly larger numbers of photons will likely rely on the continuous development of multiplexed parametric SPSs [49, 50] and de-multiplexed QD-SPSs [56, 57]. They both require a network of fast and low-loss optical switches. Recent progress in thin-film LN-based switches [118] has shown a great promise for the multiplexed and de-multiplexed many-photon generation. These advanced systems also promise fast manipulations of photons for quantum information processing and communications. Moreover, a fully integrated quantum chip having sources, circuits and detectors is yet to be realised. Much effort is being put into this, e.g., demonstrations of an integration of circuits with SNSPDs (**Fig.2j**) [30, 61], and SPSs with SNSPDs (**Fig.2h**) [119, 120]. A few exciting technical challenges need to be tackled towards full integrations, such as how to manipulate photons at cryogenic temperature required for the detector operations. Nevertheless, fully integrated QKD chips are readily achievable in the InP [37, 70] and Si/III-V platforms [71], promising miniaturized, low-costs devices allow for both quantum and classical communications.

In the last decade, integrated quantum photonics has steadily become a versatile platform that is proving invaluable in the development of future quantum technologies and applications. Through embracing wafer-scale fabrication processes, we fully anticipated that over the next decade large-scale integrated photonic devices and systems will be continuously developed to facilitating the revolutionary development of quantum communication, quantum information processing, and quantum simulation.

**Box 1: on-chip encoding quantum information in single photons**

Photons can encode quantum information in binary and *d*-ary manner. That means *qubit* or *qudit* state is represented as a superposition of eigenmodes $\sum_{i=0}^{d-1} c_i |i\rangle$, where $c_i$ is the complex amplitude in $|\lambda\rangle$ mode and *d* refers to the dimensionality. Photons own rich DoFs, while IQP offers unique abilities to on-chip control these DoFs, including polarisation, spatial mode, temporal mode, frequency, and location (path) of single photons. For example, the transverse electric $|TE\rangle$ and transverse magnetic $|TM\rangle$ modes in waveguides function a pair of polarised eigenstates. Engineering the birefringence enables the rotation of polarised qubit; its arbitrary operation presents challenging in planar optical waveguides but more controllable in circular waveguides [20]. Photons that exist in a series of time-bin $|\tau\rangle$ can be initialised and measured by integrated Franson interferometers [28], while photons that exist in a parallel of frequency-bin $|\lambda\rangle$ can be naturally produced in an optical resonator and arbitrarily manipulated with fiber-components [43]. These temporal and spectral qubits/qudits are well suited for quantum communications in optical-fiber, which are compatible with present telecommunication infrastructure. Photons that simultaneously locate at dual/multiple waveguides $|k\rangle$, forming path-encoded qubit/qudit states. These states can be arbitrarily and precisely prepared, manipulated and measured using programmable Mach-Zehnder interferometers (MZIs) [44]. Moreover, a multimode waveguide that supports more than one eigenmodes enables a new DoF to encode quantum states in the high-order spatial modes $|m\rangle$ [45]. Crucially, the multi-dimensional nature of $|\tau, \lambda, k, m\rangle$ represents a distinguishing characteristic of photons. On-chip encoding and control of multi-photon, multi-dimensional, and multi-DoFs entangled states allow a significant enhancement of quantum information processing and communication capacities with integrated photonics.

**Box 2: integrated building blocks for operating photonic states**

Quantum interference plays a key role in photonic quantum technology. Two-photon quantum interference occurs when two indistinguishable photons meet at a balanced beamsplitter (BS), and the probability of both photons being reflected or transmitted cancel out. Inducing distinguishability no matter in which DoF yields the observation of Hong-Ou-Mandel (HOM) dip or interference. We take the path-encoded qubits as an example to illustrate on-chip building blocks for quantum applications.

Directional coupler (DC) and multimode interferometer (MMI) are two types of integrated beamsplitters, whose unitary is represented as a Hadamard-like operator. A MZI having two DCs/MMIs and one re-configurable phase-shifter can reliably perform classical and quantum interference. The relative phase is induced by changing the refractive index of one arm, e.g., by thermo-optic or electro-optic effect. A combination of one MZI and two additional phases enables an arbitrary SU(2) unitary transformation. It is such a simple device that allows arbitrary, but precise preparation, operation and analysis of a path-encoded single-qubit as $\alpha |0\rangle + \beta |1\rangle$, where $|0\rangle$ and $|1\rangle$ are logical states. When the two identical photons meet at a MZI, photons at outputs present a superposition of bunching state and anti-bunching state, dependent on the phase in the MZI. This is referred to the HOM-like or reverse-HOM interference [25]. Photonic two-qubit entangling operations are enabled by quantum interference [9]. The circuit diagrams illustrate three types of two-qubit operations for path-encoded qubits: $\hat{O}_{BS}$ functions in analogy to a BS on two polarised qubits, where each qubit has a 50% probability of being found in either 1' or 2' regardless of its input; $\hat{O}_{PBS}$ transmits $|0\rangle$ and swaps $|1\rangle$ mode in analogy to a polarisation beamsplitter (PBS) transformation; $\hat{O}_{CZ}$ presents the KLM CZ operation, consisting of several BSs with different reflectivity. Let us briefly summary the basis of these operations without going details: they all allow to measuring qubits in the Bell basis, performing Bell analysis;

$\hat{O}_{PBS}$ enables the fusion operation that can be adopted to create multi-photon star-graph states; performing $\hat{O}_{CZ}$ allows the generation of cluster states. The non-interaction nature of single-photons leads to non-deterministic two-qubit operations, having a successful probability of 1/2 for $\hat{O}_{BS}$ and $\hat{O}_{PBS}$, and 1/9 for $\hat{O}_{CZ}$. Notably, the two-photon quantum interference occurs at a 2×2 BS represents the most basic scenario, and can be further generalised to *n*-photon quantum interference in a *m*-mode unitary $\hat{O}_{LO}$. A well-known example is the Boson sampling problem [13], which is believed to be an intractable task for classical computers but efficiently executable in a specific quantum machine.

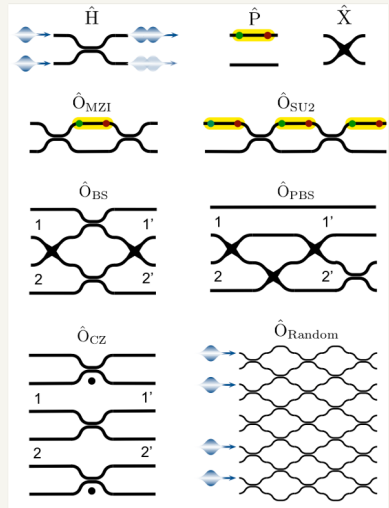

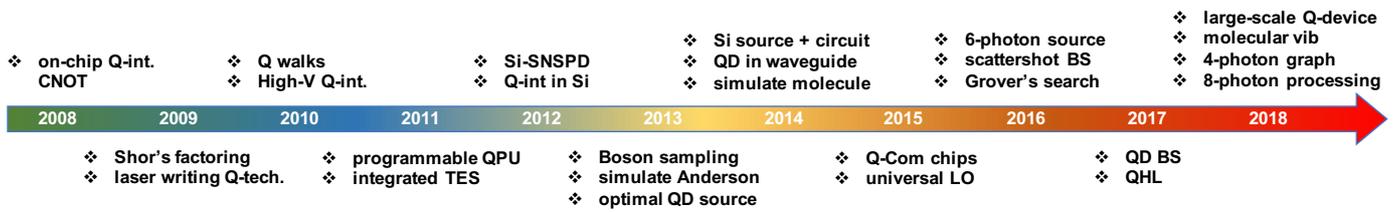

Fig. 1 Key demonstrations in integrated quantum photonics: first demonstration of on-chip quantum interference and integrated CNOT gate [14], test of Shor's factoring algorithm [76], laser-writing integrated quantum photonic circuits [19], quantum walks with correlated photons [107], near-optimal two-photon quantum interference [18], the first re-programmable multi-functional quantum processor unit (QPU) [17], integrated TES PNR detector on waveguide [62], Si-waveguide integrated SPSND with high efficiency [26], observation of quantum interference in Si [24], demonstrations of Boson sampling with photons [91–95], simulation of Anderson localizations with entangled photons [104], near-optimal single-photons generation from a single QD [52], integration of SFWM-SPSs with quantum circuits [25], high-efficiency QD source in waveguide [55], quantum simulation of molecular ground-state [79], demonstrations of chip-to-chip quantum key distribution and entanglement distribution [37, 73], demonstration of universal linear optic circuit [77], on-chip generation of 6 photon [47], first demonstration of scattershot Boson sampling [97], test of Grover's search algorithm [85], high-efficiency QD Boson sampling using a QD source [57], test of quantum Hamiltonian learning (QHL) [115], large-scale quantum circuits in Si with 670 components [44], simulation of molecular vibrations [82], demonstration of 4-photon graph state [87], and on-chip generation of 8-photon in Si [98]. The chronological order refers to the first date appeared in the public domain including conference and arXiv.

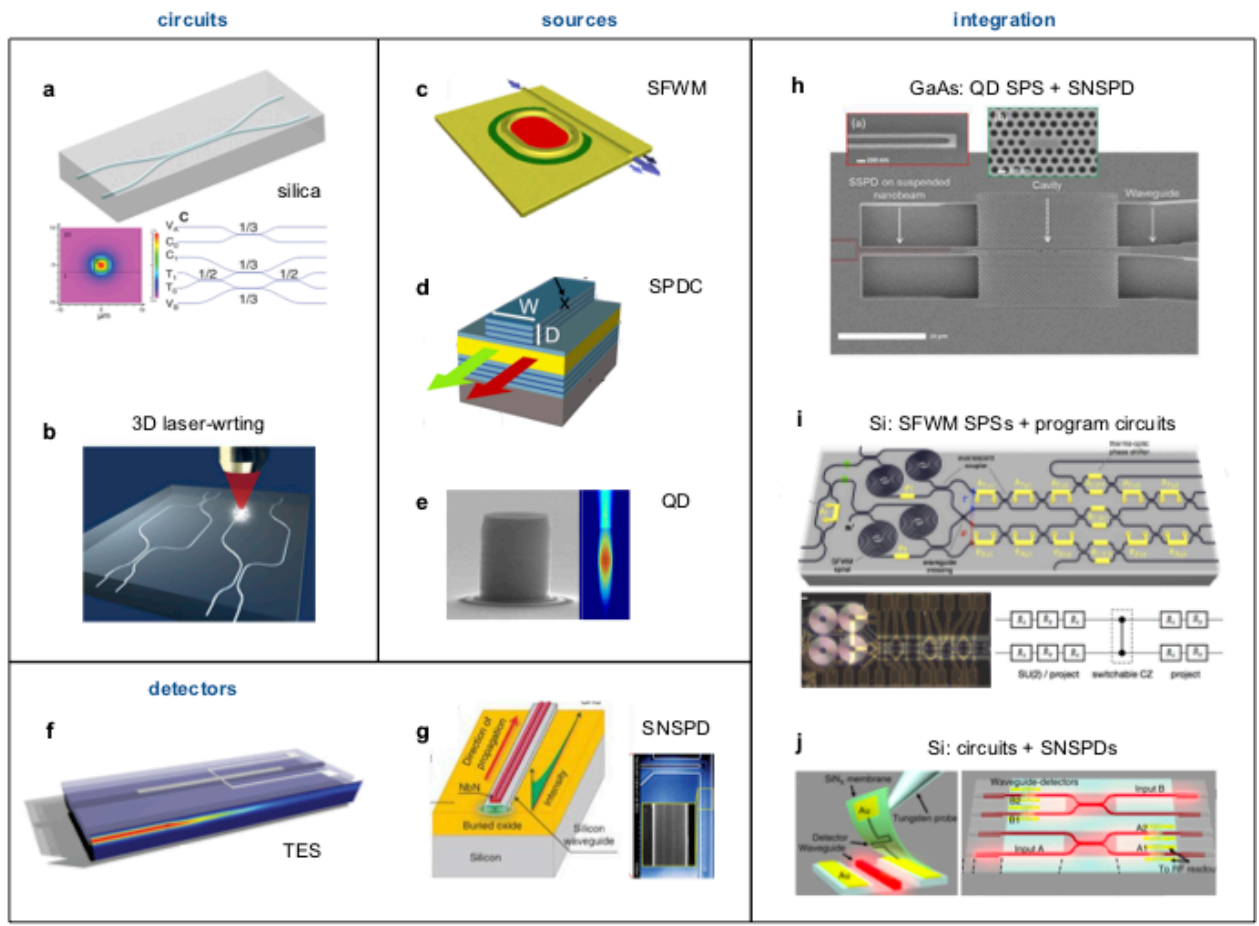

Fig. 2 Integrated single-photon sources, detectors and quantum circuits. a, the first IQP circuit for CNOT entangling gate in silica-on-insulator waveguides. b, laser-writing quantum circuit in silica with complex 3D geometry. c, an integrated SFWM-SPS in Si-photonic waveguide. Engineering microresonator enables the generation of highly pure and indistinguishable photons. d, an integrated SPDC-SPS in GaAs waveguides. A Bragg-waveguide that sustains two different types of eigenmodes are engineered to meet the phase matching condition. e, a semiconductor InGaAs/GaAs QD SPS able to emit near-deterministic and highly pure single-photons from a single dot embedded in a micropillar. f, an integrated TES SPD that is evanescently coupled to silica waveguides. Up to 5 photons guided in the waveguides can be resolved by the integrated TES. g, an integrated SNSPD atop of Si-waveguide, that can absorb and detect photons with >90% efficiency. h, an integration of a QD-SPS with SNSPDs in the GaAs photonic-crystal waveguide system. It enables the detection of QD luminescence in a compact system. i, an integration of SFWM-SPSs with re-programmable quantum photonic circuits in Si-photonics. It allows the on-chip preparation, manipulation and measurement of photonic states. j, an integration of Si-photonic circuits with SNSPDs by using the μm-scale flip-chip process. It allows on-chip measurements of non-classical light. Figure reproduced from: a, ref. [14], AAAS; b, ref. [80], Macmillan Publishers Ltd; c, ref. [46], OSA; d, ref. [34], APS; e, ref. [53], APS; f, ref. [62], APS; g, ref. [26], Macmillan Publishers Ltd; h, ref. [118], MDPI; i, ref. [39], IOPscience; j, ref. [58], Macmillan Publishers Ltd.

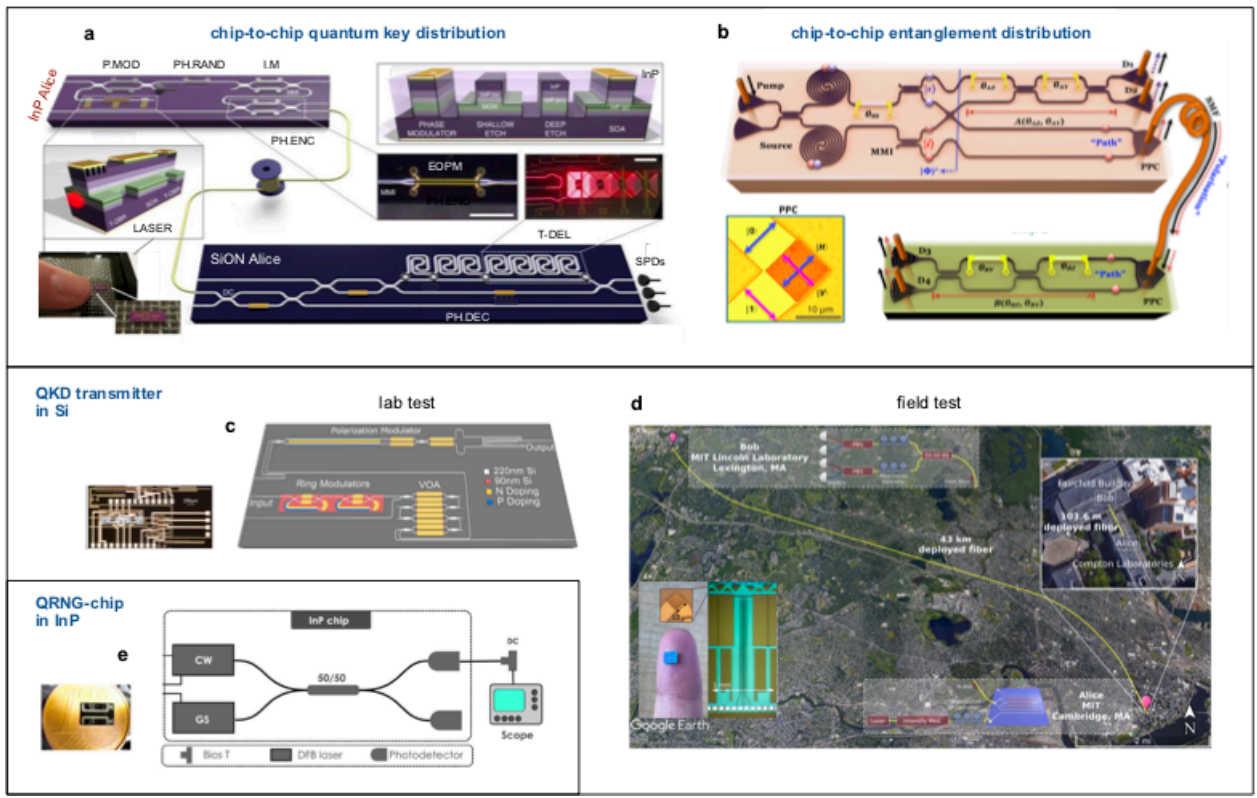

**Fig. 3 Integrated chip-based quantum communications. a,** a chip-to-chip QKD system between a 2×6 mm$^2$ InP transmitter and a 2×32 mm$^2$ SiO$_x$N$_y$ receiver. The InP transmitter consists of a tunable laser, electro-optic phase modulators (EOPMs) and MZIs, allowing for pulse modulation (P.MOD), phase randomization (PH.RAND) and intensity modulation (I.M). Qubits are prepared by phase encoding (PH.ENC). The SiO$_x$N$_y$ receiver consists of several thermal-optic phase shifters (TOPSs) and reconfigurable delay lines, allowing for phase decoding (PH.DEC). **b,** a chip-to-chip entanglement distribution system between two Si-photonic chips linked by a 10m single-mode fiber. Path-entangled states are generated on a 1.2×0.5 mm$^2$ chip with two SFWM-SPSs, and then coherently distributed to a 0.3×0.05 mm$^2$ chip, by using polarisation-path-converters (PPCs) based on the 2D grating couplers. **c,** a Si-based transmitter for polarisation-encoded QKD. The transmitter consists of ring modulator, attenuator and polarisation modulator, in a 1.3×3 mm$^2$ area. The ring modulators generate periodic ns pulse trains, while the polarisation modulator prepares two sets of orthogonal polarized states for BB84 QKD. **d,** an intercity metropolitan QKD system based on a Si transmitter. Alice is located in Cambridge and Bob in Lexington, connected by a 43-km dark fiber. Alice prepares the polarisation states using a 10 Gbps EOPM and a PPC, while Bob measures the states using a fiber-optic receiver. **e,** a fully integrated QRNG in an InP chip. The device consists of two distributed feedback lasers, a balanced MMI beamsplitter and two photodetectors for heterodyne detection. Figure reproduced from: **a,** ref. [37] , Macmillan Publishers Ltd; **b,** ref. [73], OSA; **c,** ref. [65], OSA; **d,** ref. [67], APS; **e,** ref. [38], OSA.

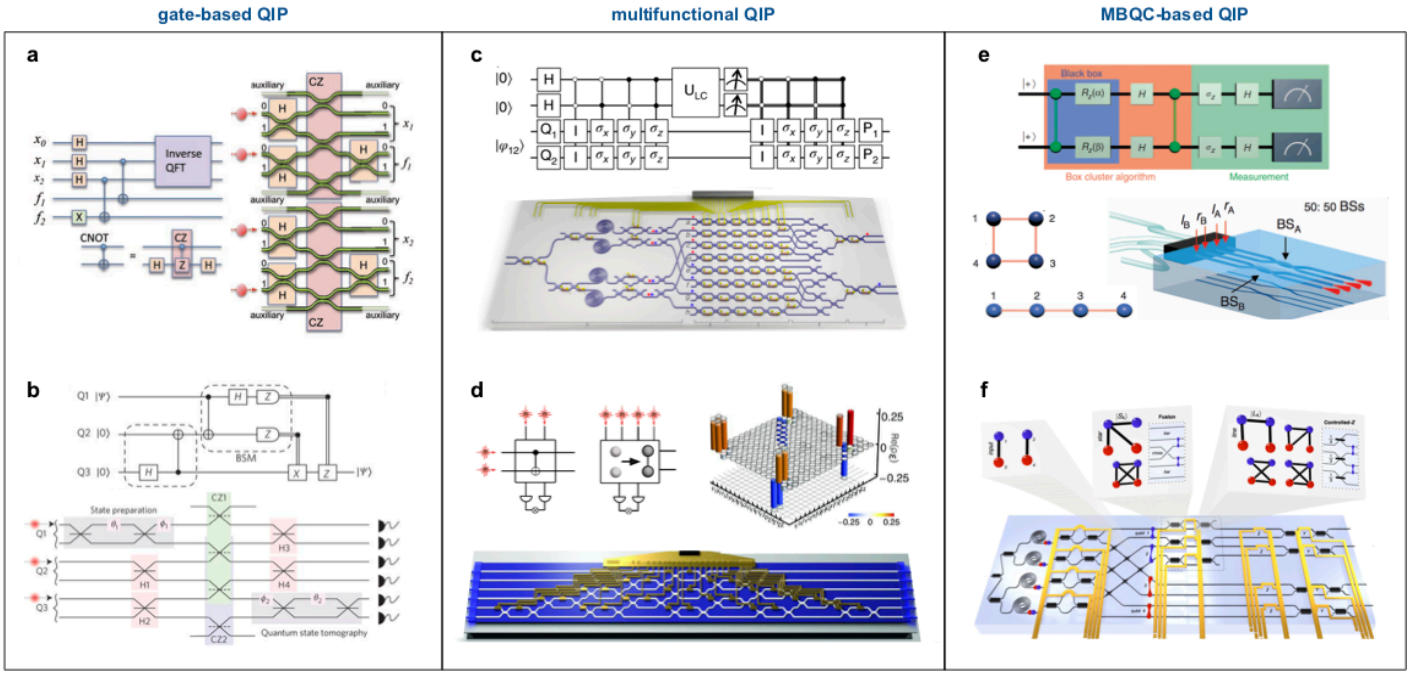

**Fig. 4 Quantum information processing and quantum computing with integrated photonics. a,** and **b,** implementations of gate-based quantum information processing (QIP). In **a,** a compiled version of Shor's factoring algorithm is performed in a SiO2 chip that consists of 2 CZ gates and 6 Hadamard gates. Four photons from bulk SPDC-SPSs were coupled into the photonic chip. In **b,** quantum teleportation of photonic qubits is performed in a SiO2 chip. An entangled state encoded on qubits [Q2, Q3] are prepared, and a Bell state measurement is performed on [Q1, Q2] qubits. **c,** and **d,** multi-functional QIP using fully re-programmable quantum photonic chips. In **c,** a universal two-qubit quantum processor is realised in a re-programmable Si-chip, which monolithically integrates 4 SFWM-SPSs, 58 TOPSs and 82 MMIs. The device was re-programmed to implement 98 two-qubit operations with high fidelities, and multiple algorithms were implemented. In **d,** a six-mode universal linear-optic circuit is realised in a re-programmable SiO2 chip. The device consists of a cascade of 15 MZIs and 30 TOPSs, and up to 6 photons were injected into the chip from bulk-optic SPDC-SPSs. The device was programmed to implement heralded CNOT gate, Boson sampling, 6-dimensional complex Hadamards, and simulations of molecular dynamics. **e,** and **f,** implementations of MBQC in integrated photonic chips. In **e,** four-qubit linear- and box-cluster states are generated in a laser-writing photonic chip. This is enabled by on-chip manipulating the two-photon four-qubit hyperentangled state and cluster state, that are encoded in photon's path and polarisation DoFs. The Grover's search algorithm on a four-element database was implemented by controlling the device. In **f,** four-photon four-qubit star- and line-graph states are generated, manipulated and measured in a Si-photonic chip. Four photons are generated in 4 waveguide SFWM-SPSs. Qubits are operated on a reconfigurable entangling gate, performing either a fusion operation or CZ operation. Figure reproduced from: **a,** ref. [76], AAAS; **b,** ref. [78], Macmillan Publishers Ltd; **c,** ref. [81], Macmillan Publishers Ltd; **d,** ref. [77], AAAS; **e,** ref. [85], Macmillan Publishers Ltd;. **f,** ref. [87].

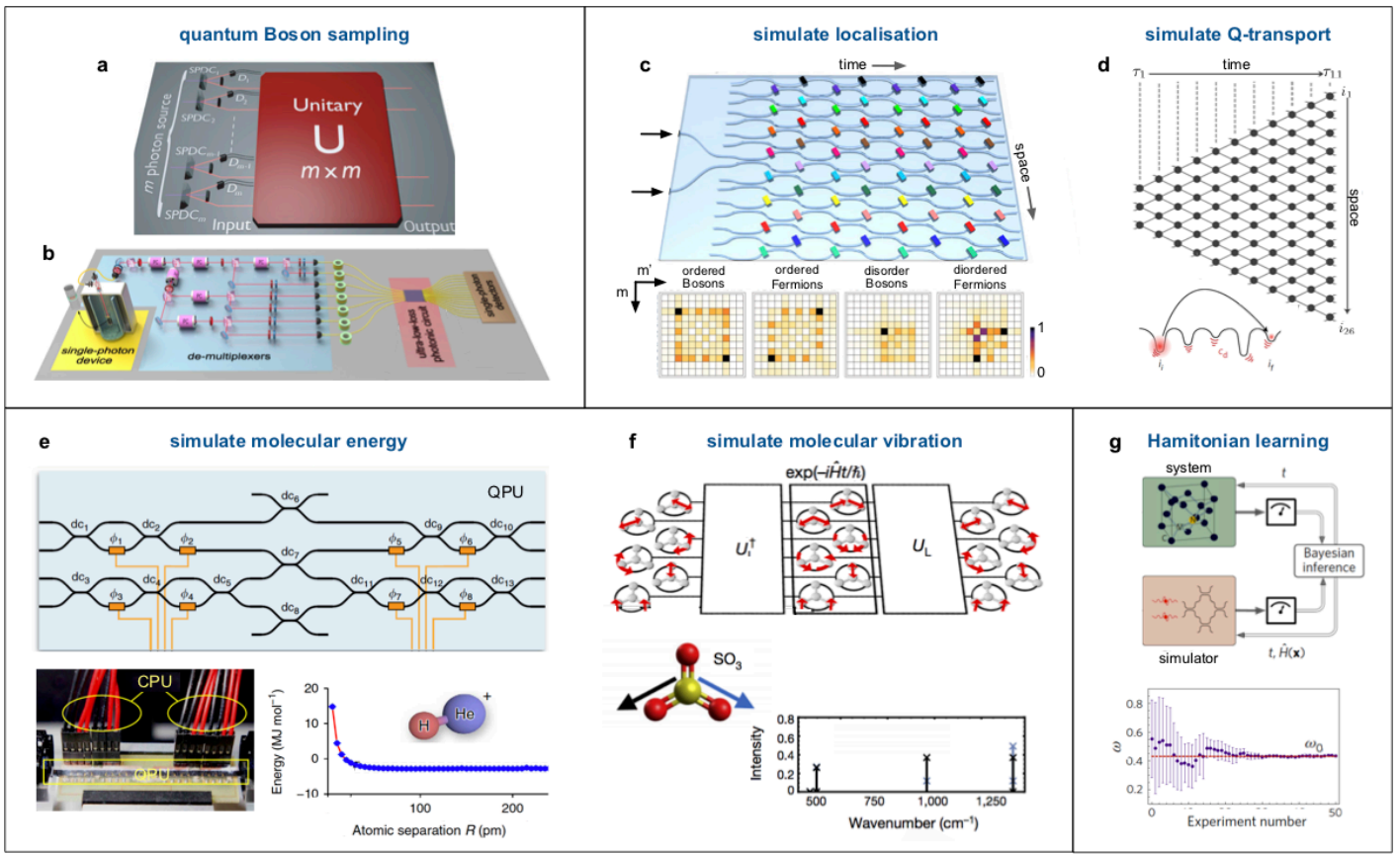

**Fig. 5 On-chip Boson sampling and quantum simulation with photons. a,** and **b,** scattershot Boson sampling and quantum dot-based Boson sampling. In **a,** the first scattershot Boson sampling with 3-photon is implemented using 6 bulk SPDC-SPSs in a 13-mode laser-writing SiO$_2$ chip. Scattershot can reach an exponential enhancement in sampling rate, by simultaneously pumping $k$ parametric sources, and measuring $n$ photons in a random but heralded set of input ports ($k \gg n$). **b,** implementation of 5-photon QD Boson sampling with a high rate of 4 Hz. Near-optimal single-photons emitted from a single QD-SPS are actively de-multiplexed into 5 spatial modes, and the 5 photons are scattered in a ultra-low-loss 9×9 photonic circuit. **c,** and **d,** simulations of physical quantum phenomena and processes via quantum walks. In **c,** Anderson localization is observed with two-particles in laser-writing chip, that is able to perform 1D QW with 16 spatial steps and up to 8 time steps. Symmetry/anti-symmetry polarisation-entangled states were injected into the circuits to mimic the behavior of bosons and fermions. In **d,** environment-assisted QT is observed with single-particle (coherent light) by controlling and mapping static and dynamic disorders in a re-programmable Si device. The device consists of 176 TOPSs and 88 MZIs in a 4.9×2.4mm$^2$ area, and it enables ultra-high precision controls of photons. **e,** and **f,** simulations of ground-state eigenenergy and vibrionic dynamics of molecules using re-programmable quantum chips. In **e,** a quantum variational eigensolver algorithm is implemented by combining a SiO$_2$ photonic quantum processor with a classical computer. It allows the estimations of ground-state energy and bond dissociation curve for the He-H$^+$ molecule. In **f,** exploiting a natural mapping between vibrations in molecules and photons in waveguides allows the simulation of vibrational dynamics of the atoms within molecule. Several 4-atom molecules with up to 6-mode vibrational modes are simulated using the 6-mode programable SiO$_2$ chip. **g,** an interface between a quantum simulator and target quantum system is enabled by a quantum Hamiltonian learning approach. A programmable Si-photonics quantum simulator and an electron spin system in a nitrogen-vacancy center in diamond was interfaced, and the former was used to simulate the dynamics of the latter. Figure reproduced from: **a,** ref. [97], AAAS; **b,** ref. [57, 58], APS; **c,** ref. [105], Macmillan Publishers Ltd; **d,** ref. [41], Macmillan Publishers Ltd; **e,** ref. [79], Macmillan Publishers Ltd; **f,** ref. [82], Macmillan Publishers Ltd; **g,** ref. [115], Macmillan Publishers Ltd.

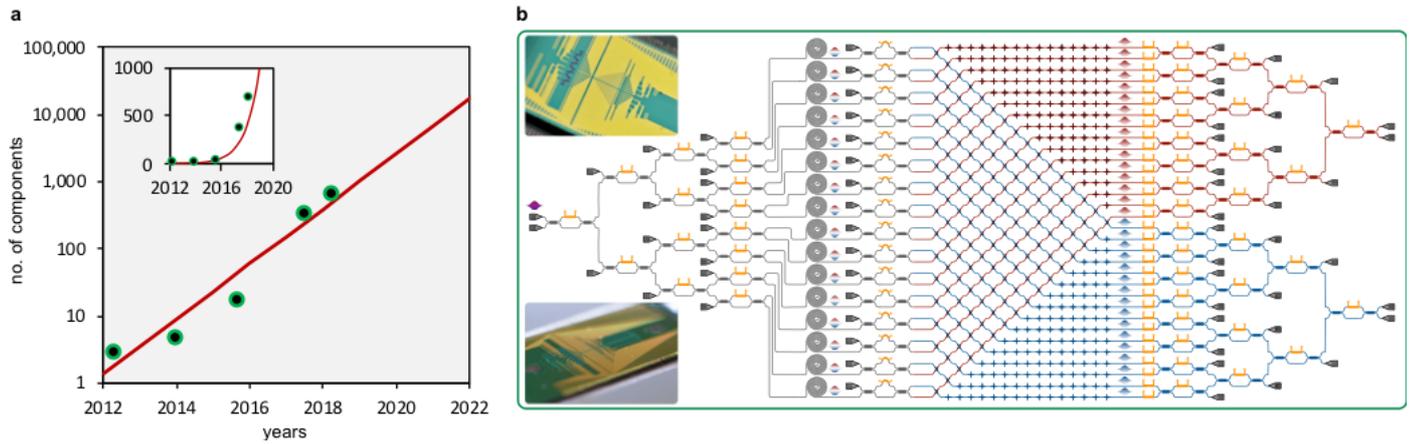

**Fig. 6 Towards large-scale quantum photonic circuits. a,** scaling of the number of quantum photonic components that are monolithically integrated in a single silicon chip: from the first on-chip HOM quantum interference experiment in silicon with an adoption of a single MZI in 2012 [24], to the state-of-the-art large-scale integrated quantum photonic circuit in 2018, as shown in **b.** This large-scale device monolithically integrates 16 SFMW-SPSs, 93 reconfigurable TOPSs, 122 MMIs, 376 waveguide-crossers and 64 grating couplers, in total 671 photonic components. The silicon device was fabricated using the standard CMOS process. It allows the fully on-chip generation, manipulation and measurement of multi-dimensional entanglement with high controllability and universality, and up to 15×15 dimensional two-photon entangled states have been verified. Figure reproduced from: **b,** ref. [44], AAAS.

# References

1. S. J. Freedman & J. F. Clauser. Experimental test of local hidden-variable theories. *Phys. Rev. Lett.* **28,** 938–941 (1972).
2. L.-A. Wu, H. J. Kimble, J. L. Hall & H. Wu. Generation of squeezed states by parametric down conversion. *Phys. Rev. Lett.* **57,** 2520–2523 (1986).
3. D. Bouwmeester *et al.* Experimental quantum teleportation. *Nature* **390,** 575–579 (1997).
4. L. K. Shalm *et al.* Strong loophole-free test of local realism. *Phys. Rev. Lett.* **115,** 250402 (2015).
5. N. Gisin, G. Ribordy, W. Tittel & H. Zbinden. Quantum cryptography. *Rev. Mod. Phys.* **74,** 145–195 (2002).
6. T. D. Ladd *et al.* Quantum computers. *Nature* **464,** 45–53 (2010).
7. A. Aspuru-Guzik & P. Walther. Photonic quantum simulators. *Nat. Phys.* **8,** 285–291 (2012).
8. D. D. Awschalom, R. Hanson, J. Wrachtrup & B. B. Zhou. Quantum technologies with optically interfaced solid-state spins. *Nat. Photon.* **12,** 516–527 (2018).
9. E. Knill, R. Laflamme & G. J. Milburn. A scheme for efficient quantum computation with linear optics. *Nature* **409,** 46–52 (2000).
10. D. Gottesman & I. L. Chuang. Demonstrating the viability of universal quantum computation using teleportation and single-qubit operations. *Nature* **402,** 390–393 (1999).
11. R. Raussendorf & H. J. Briegel. A one-way quantum computer. *Phys. Rev. Lett.* **86,** 5188–5191 (2001).
12. M. A. Nielsen. Optical quantum computation using cluster states. *Phys. Rev. Lett.* **93,** 040503 (2004).
13. S. Aaronson & A. Arkhipov. *The Computational Complexity of Linear Optics* in *Proceedings of the Forty-third Annual ACM Symposium on Theory of Computing* (2011), 333–342.
14. A. Politi, M. J. Cryan, J. G. Rarity, S. Yu & J. L. O'Brien. Silica-on-silicon waveguide quantum circuits. *Science* **320,** 646–649 (2008).
15. H.-S. Zhong *et al.* 12-photon entanglement and scalable scattershot Boson sampling with optimal entangled-photon pairs from parametric down-conversion. *Phys. Rev. Lett.* **121,** 250505 (2018).
16. J. C. F. Matthews, A. Politi, S. Andre & J. L. O'Brien. Manipulation of multiphoton entanglement in waveguide quantum circuits. *Nat. Photon.* **3,** 346–350 (2009).
17. P. J. Shadbolt *et al.* Generating, manipulating and measuring entanglement and mixture with a reconfigurable photonic circuit. *Nat. Photon.* **6,** 45–49 (2012).
18. A. Laing *et al.* High-fidelity operation of quantum photonic circuits. *Appl. Phys. Lett.* **97,** 211109 (2010).
19. B. J. Smith, D. Kundys, N. Thomas-Peter, P. G. R. Smith & I. A. Walmsley. Phase-controlled integrated photonic quantum circuits. *Opt. Express* **17,** 13516–13525 (2009).
20. G. Corrielli *et al.* Rotated waveplates in integrated waveguide optics. *Nat. Commun.* **5,** 4249 (2014).
21. L. Sansoni *et al.* Polarization entangled state measurement on a chip. *Phys. Rev. Lett.* **105,** 200503 (2010).
22. A. Crespi *et al.* Integrated photonic quantum gates for polarization qubits. *Nat. Commun.* **2,** 566 (2011).
23. H. Takesue *et al.* Entanglement generation using silicon wire waveguide. *Appl. Phys. Lett.* **91,** 201108 (2007).
24. D. Bonneau *et al.* Quantum interference and manipulation of entanglement in silicon wire waveguide quantum circuits. *New J. Phys.* **14,** 045003 (2012).
25. J. W. Silverstone *et al.* On-chip quantum interference between silicon photon-pair sources. *Nat. Photon.* **8,** 104–108 (2013).
26. W. H. P. Pernice *et al.* High-speed and high-efficiency travelling wave single-photon detectors embedded in nanophotonic circuits. *Nat. Commun.* **3,** 1325–10 (2012).
27. X. Lu *et al.* Chip-integrated visible-telecom entangled photon pair source for quantum communication. *Nat. Phys.* **15,** 373–381 (2019).
28. X. Zhang *et al.* Integrated silicon nitride time-bin entanglement circuits. *Opt. Lett.* **43,** 3469–3472 (2018).
29. A. Dutt *et al.* On-chip optical squeezing. *Phys. Rev. Applied* **3,** 044005 (2015).
30. C. Schuck *et al.* Quantum interference in heterogeneous superconducting-photonic circuits on a silicon chip. *Nat. Commun.* **7,** 10352 (2016).
31. S. Tanzilli *et al.* PPLN waveguide for quantum communication. *Eur. Phys. J. D - Atomic, Molecular, Optical and Plasma Physics* **18,** 155–160 (2002).
32. H. Jin *et al.* On-chip generation and manipulation of entangled photons based on reconfigurable Lithium-Niobate waveguide circuits. *Phys. Rev. Lett.* **113,** 103601 (2014).
33. J. P. Höpker *et al.* Towards integrated superconducting detectors on lithium niobate waveguides. *arXiv,* 1708.06232 (2017).
34. R. Horn *et al.* Monolithic source of photon pairs. *Phys. Rev. Lett.* **108,** 153605 (2012).
35. J. Wang *et al.* Gallium arsenide (GaAs) quantum photonic waveguide circuits. *Opt. Commun.* **327,** 49–55 (2014).
36. J. P. Sprengers *et al.* Waveguide superconducting single-photon detectors for integrated quantum photonic circuits. *Appl. Phys. Lett.* **99,** 181110 (2011).
37. P. Sibson *et al.* Chip-based quantum key distribution. *Nat. Commun.* **8,** 13984 (2017).
38. C. Abellan *et al.* Quantum entropy source on an InP photonic integrated circuit for random number generation. *Optica* **3,** 989–994 (2016).
39. R Santagati *et al.* Silicon photonic processor of two-qubit entangling quantum logic. *J. Opt.* **19,** 114006 (2017).
40. F. Caruso, A. Crespi, A. G. Ciriolo, F. Sciarrino & R. Osellame. Fast escape of a quantum walker from an integrated photonic maze. *Nat. Commun.* **7,** 11682 (2016).
41. N. C. Harris *et al.* Quantum transport simulations in a programmable nanophotonic processor. *Nat. Photon.* **11,** 447–452 (2017).
42. T. Rudolph. Why I am optimistic about the silicon-photonic route to quantum computing. *APL Photonics* **2,** 030901 (2017).
43. J. M. Lukens & P. Lougovski. Frequency-encoded photonic qubits for scalable quantum information processing. *Optica* **4,** 8–16 (2017).
44. J. Wang *et al.* Multidimensional quantum entanglement with large-scale integrated optics. *Science* **360,** 285–291 (2018).
45. L.-T. Feng *et al.* On-chip coherent conversion of photonic quantum entanglement between different degrees of freedom. *Nat. Commun.* **7,** 11985 (2016).
46. E. Engin *et al.* Photon pair generation in a silicon micro-ring resonator with reverse bias enhancement. *Opt. Express* **21,** 27826–27834 (2013).
47. J. B. Spring *et al.* Chip-based array of near-identical, pure, heralded single-photon sources. *Optica* **4,** 90–7 (2017).
48. D. Llewellyn *et al.* Demonstration of chip-to-chip quantum teleportation in *Conference on Lasers Electro-Optics (CLEO) 2019* (OSA, 2019), JTh5C.4.
49. F. Kaneda & K. Paul. High-efficiency single-photon generation via large-scale active time multiplexing. *arXiv,* 1803.04803 (2017).
50. M. J. Collins *et al.* Integrated spatial multiplexing of heralded single-photon sources. *Nat. Commun.* **4,** 2582 (2014).
51. P. Senellart, G. Solomon & A. White. High-performance semiconductor quantum-dot single-photon sources. *Nat. Nanotech.* **12,** 1026–1039 (2017).
52. Y.-M. He *et al.* On-demand semiconductor single-photon source with near-unity indistinguishability. *Nat. Nanotech.* **8,** 213–217 (2013).
53. X. Ding *et al.* On-demand single photons with high extraction efficiency and near-unity indistinguishability from a resonantly driven quantum dot in a micropillar. *Phys. Rev. Lett.* **116,** 020401 (2016).
54. N. Somaschi *et al.* Near-optimal single-photon sources in the solid state. *Nat. Photon.* **10,** 340–345 (2016).
55. M. Arcari *et al.* Near-unity coupling efficiency of a quantum emitter to a photonic crystal waveguide. *Phys. Rev. Lett.* **113,** 093603 (2014).
56. F. Lenzini *et al.* Active demultiplexing of single photons from a solid-state source. *Laser Photonics Rev.* **11,** 1600297 (2017).
57. H. Wang *et al.* High-efficiency multiphoton boson sampling. *Nat. Photon.* **11,** 361–365 (2017).

58. H. W. et al. Toward scalable boson sampling with photon loss. *Phys. Rev. Lett.* **120,** 230502 (2018).
59. J. C. Loredo *et al.* Boson sampling with single-photon Fock states from a bright solid-state source. *Phys. Rev. Lett.* **118,** 130503 (2017).
60. R. H. Hadfield. Single-photon detectors for optical quantum information applications. *Nat. Photon.* **3,** 696–705 (2009).
61. F. Najafi *et al.* On-chip detection of non-classical light by scalable integration of single-photon detectors. *Nat. Commun.* **6** (2015).
62. T. Gerrits *et al.* On-chip, photon-number-resolving, telecommunication-band detectors for scalable photonic information processing. *Phys. Rev. A* **84,** 060301 (2011).
63. D. Sahin *et al.* Waveguide photon-number-resolving detectors for quantum photonic integrated circuits. *Appl. Phys. Lett.* **103,** 111116 (2013).
64. T. Honjo, K. Inoue & H. Takahashi. Differential-phase-shift quantum key distribution experiment with a planar light-wave circuit Mach–Zehnder interferometer. *Opt. Lett.* **29,** 2797–2799 (2004).
65. C. Ma *et al.* Silicon photonic transmitter for polarization-encoded quantum key distribution. *Optica* **3,** 1274–1278 (2016).
66. P. Sibson *et al.* Integrated silicon photonics for high-speed quantum key distribution. *Optica* **4,** 172–177 (2017).
67. D. Bunandar *et al.* Metropolitan quantum key distribution with silicon photonics. *Phys. Rev. X* **8,** 021009 (2018).
68. M. G. Thompson. *Large-scale integrated quantum photonic technologies for communications and computation* in *Optical Fiber Communication Conference (OFC) 2019* (OSA, 2019), W3D.3.
69. Y. Ding *et al.* High-dimensional quantum key distribution based on multicore fiber using silicon photonic integrated circuits. *npj Quantum Information* **3,** 25 (2017).
70. H. Semenenko, P. Sibson, M. G. Thompson & C. Erven. Interference between independent photonic integrated devices for quantum key distribution. *Opt. Lett.* **44,** 275–278 (2019).
71. C. Agnesi *et al.* Hong–Ou–Mandel interference between independent III–V on silicon waveguide integrated lasers. *Opt. Lett.* **44,** 271–274 (2019).
72. C. Autebert *et al.* Integrated AlGaAs source of highly indistinguishable and energy-time entangled photons. *Optica* **3,** 143–146 (2016).
73. J. Wang *et al.* Chip-to-chip quantum photonic interconnect by path-polarization interconversion. *Optica* **3,** 407–413 (2016).
74. T. Roger *et al.* Real-time interferometric quantum random number generation on chip. *J. Opt. Soc. Am. B* **36,** B137–B142 (2019).
75. F. Raffaelli *et al.* A homodyne detector integrated onto a photonic chip for measuring quantum states and generating random numbers. *Quantum Science and Technology* **3,** 025003 (2018).
76. A. Politi, J. C. F. Matthews & J. L. O'Brien. Shor's quantum factoring algorithm on a photonic chip. *Science* **325,** 1221 (2009).
77. J. Carolan *et al.* Universal linear optics. *Science* **349,** 711–716 (2015).
78. B. J. Metcalf *et al.* Quantum teleportation on a photonic chip. *Nat. Photon.* **8,** 770–774 (2014).
79. A. Peruzzo *et al.* A variational eigenvalue solver on a photonic quantum processor. *Nat. Commun.* **5,** 4213 (2013).
80. F. Flamini *et al.* Thermally reconfigurable quantum photonic circuits at telecom wavelength by femtosecond laser micromachining. *Light Sci. Appl.* **4,** e354 (2015).
81. X. Qiang *et al.* Large-scale silicon quantum photonics implementing arbitrary two-qubit processing. *Nat. Photon.* **12,** 534–539 (2018).
82. C. Sparrow *et al.* Simulating the vibrational quantum dynamics of molecules using photonics. *Nature* **557,** 660–667 (2018).
83. C. Taballione *et al.* 8x8 Reconfigurable Quantum Photonic Processor based on Silicon Nitride Waveguides. *arXiv,* 1805.10999 (2018).
84. J. W. Silverstone *et al.* Qubit entanglement between ring-resonator photon-pair sources on a silicon chip. *Nat. Commun.* **6,** 7948 (2015).
85. M. A. Ciampini *et al.* Path-polarization hyperentangled and cluster states of photons on a chip. *Light Sci. Appl.* **5,** e16064 (2016).
86. C. Reimer *et al.* High-dimensional one-way quantum processing implemented on d-level cluster states. *Nat. Phys.* **15,** 148–153 (2019).
87. J. C. Adcock, C. Vigliar, R. Santagati, J. W. Silverstone & M. G. Thompson. Programmable four-photon graph states on a silicon chip. *arXiv,* 1811.03023 (2018).
88. A. W. Harrow & A. Montanaro. Quantum computational supremacy. *Nature* **549,** 203–209 (2017).
89. N. Spagnolo *et al.* Three-photon bosonic coalescence in an integrated tritter. *Nat. Commun.* **4,** 1606 (2013).
90. B. J. Metcalf *et al.* Multiphoton quantum interference in a multiport integrated photonic device. *Nat. Commun.* **4,** 1356 (2013).
91. A. Crespi *et al.* Integrated multimode interferometers with arbitrary designs for photonic boson sampling. *Nat. Photon.* **7,** 545 (2013).
92. M. Tillmann *et al.* Experimental boson sampling. *Nat. Photon.* (2013).
93. J. B. Spring *et al.* Boson sampling on a photonic chip. *Science* **339,** 798–801 (2013).
94. J. Carolan *et al.* On the experimental verification of quantum complexity in linear optics. *Nat. Photon.* **8,** 621–626 (2014).
95. M. A. Broome *et al.* Photonic boson sampling in a tunable circuit. *Science* **339,** 794–798 (2013).
96. M. Tillmann *et al.* Generalized multiphoton quantum interference. *Phys. Rev. X* **5,** 041015 (4 2015).
97. M. Bentivegna *et al.* Experimental scattershot boson sampling. *Sci. Adv.* **1,** e1400255 (2015).
98. S. Paesani *et al.* Generation and sampling of quantum states of light in a silicon chip. *Arxiv,* 1812.03158 (2018).
99. S. Aaronson & A. Arkhipov. Boson sampling is far from uniform. *Quantum Inf. Comput.* **14,** 1383–423 (2014).
100. N. Spagnolo *et al.* Experimental validation of photonic boson sampling. *Nat. Photon.* **8,** 615 (2014).
101. T. Giordani *et al.* Experimental statistical signature of many-body quantum interference. *Nat. Photon.* **12,** 173–178 (2018).
102. I. Agresti *et al.* Pattern recognition techniques for Boson Sampling validation. *Phys. Rev. X* **9,** 011013 (2019).
103. A. Neville *et al.* Classical boson sampling algorithms with superior performance to near-term experiments. *Nat. Phys.* **13,** 1153–1157 (2017).
104. L. Sansoni *et al.* Two-particle bosonic-fermionic quantum walk via integrated photonics. *Phys. Rev. Lett.* **108,** 010502 (2012).
105. A. Crespi *et al.* Anderson localization of entangled photons in an integrated quantum walk. *Nat. Photon.* **7,** 322 (2013).
106. I. Pitsios *et al.* Photonic simulation of entanglement growth and engineering after a spin chain quench. *Nat. Commun.* **8,** 1569 (2017).
107. A. Peruzzo *et al.* Quantum walks of correlated photons. *Science* **329,** 1500–1503 (2010).
108. A. Crespi *et al.* Particle statistics affects quantum decay and Fano interference. *Phys. Rev. Lett.* **114,** 090201 (2015).
109. D. N. Biggerstaff *et al.* Enhancing coherent transport in a photonic network using controllable decoherence. *Nat. Commun.* **7,** 11282 (2016).
110. H. Tang *et al.* Experimental quantum fast hitting on hexagonal graphs. *Nat. Photon.* **12,** 754–758 (2018).
111. K. Poulios *et al.* Quantum walks of correlated photon pairs in two-dimensional waveguide arrays. *Phys. Rev. Lett.* **112,** 143604 (2014).
112. S. Paesani *et al.* Experimental bayesian quantum phase estimation on a silicon photonic chip. *Phys. Rev. Lett.* **118,** 100503 (2017).
113. R. Santagati *et al.* Witnessing eigenstates for quantum simulation of Hamiltonian spectra. *Sci. Adv.* **4,** eaap9646 (2018).
114. J. Huh, G. G. Guerreschi, B. Peropadre, J. R. McClean & A. Aspuru-Guzik. Boson sampling for molecular vibronic spectra. *Nat. Photon.* **9,** 615 –620 (2015).
115. J. Wang *et al.* Experimental quantum Hamiltonian learning. *Nat. Phys.* **13,** 551–555 (2017).
116. M. Raval, A. Yaacobi, and M. R. Watts. Integrated visible light phased array system for autostereoscopic image projection. *Opt. Lett.* 43, 3678-3681 (2018).

117. T. J. Seok, K. Kwon, J. Henriksson, J. Luo & M. C. Wu. *240240 Wafer-Scale Silicon Photonic Switches* in *Optical Fiber Communication Conference (OFC) 2019* (OSA, 2019), Th1E.5.
118. C. Wang *et al.* Integrated lithium niobate electro-optic modulators operating at CMOS-compatible voltages. *Nature* **562,** 101–104 (2018).
119. G. E. Digeronimo *et al.* Integration of single-photon sources and detectors on GaAs. *Photonics* **3** (2016).
120. S. Khasminskaya *et al.* Fully integrated quantum photonic circuit with an electrically driven light source. *Nat. Photon.* **10,** 727– (2016).

## Acknowledgements

J.W. acknowledges support from Beijing Academy of Quantum Information Sciences (No. Y18G21) and the Key RD Program of Guang- dong Province (Grant No. 2018B030329001). F.S. acknowledges support from the H2020-FETPROACT-2014Grant QUCHIP (Quantum Simulation on a Photonic Chip; Grant Agreement No. 641039), A.L. acknowledges support from an EPSRC Early Career Fellowship EP/N003470/1. M.G.T. acknowledges support from an ERC starter grant (ERC-2014-STG 640079) and an EPSRC Early Career Fellow- ship (EP/K033085/1).

## Author contributions

All authors contributed significantly to the preparation of the manuscript.